\documentstyle[preprint,aps]{revtex}

\draft

\begin{document}
\title{Entanglement and State Preparation}
\author{Morton H. Rubin}
\address{Department of Physics \\
University of Maryland, Baltimore County \\
Baltimore, MD 21228-5398}
\date{Dec. 22, 1998 }
\maketitle

\begin{abstract}
When a subset of particles in an entangled state is measured, the state of
the subset of unmeasured particles is determined by the outcome of the
measurement. This first measurement may be thought of as a state preparation
for the remaining particles. This type of measurement is important in
quantum computing, quantum information theory and in the preparation of
entangled states such as the Greenberger, Horne, and Zeilinger state.

In this paper, we examine how the duration of the first measurement effects
the state of the unmeasured subsystem. We discuss the case for which the
particles are photons, but the theory is sufficiently general that it can be
converted to a discussion of any type of particle. The state of the
unmeasured subsytem will be a pure or mixed state depending on the nature of
the measurement.

In the case of quantum teleportation we show that there is an eigenvalue
equation which must be satisfied for accurate teleportation. This equation
provides a limitation to the states that can be accurately teleported.
\end{abstract}


\section{Introduction}

The preparation of states of a system is one of the primitive notions in
quantum theory \cite{Peres}. It consists of a set of rules for preparing a
physical state of a given system in the laboratory and for associating a
corresponding mathematical state in the Hilbert space defined by the system.
In this paper we examine how entangled states can be used for state
preparation. This is of interest in quantum information theory, quantum
computing and in the preparation of special states such as the
Greenberger-Horne-Zeilinger (GHZ) state \cite{GHZ}. The specific question
addressed is, ``after a measurement is completed on a subset of particles in
an entangled state, what is the state of the remaining particles?'' We can
formulate this as a special case of general correlation measurements in
which one set of measurements must be completed before any further
measurements are made. That is, the first measurement or set of measurements
acts as a trigger which defines the state of the remaining particles.
Alternatively, we may use the language of probability theory and say that we
are studying a conditional amplitude of a subsystem, conditioned by the
outcome of the measurement of a second subsystem..

An interesting example of state preparation is found in quantum
teleportation \cite{teleportation}. Recall that in this case Bob and Alice
share an entangled two particle state, 
\begin{equation}
|\Psi \rangle _{AB}=\sqrt{\frac{1}{2}}\left( |+\rangle _{A}|-\rangle
_{B}-|-\rangle _{A}|+\rangle _{B}\right) ,  \label{EPRstate}
\end{equation}
and Alice is given an arbitrary state, 
\begin{equation}
|\phi \rangle _{C}=\alpha _{+}|+\rangle _{C}+\alpha _{-}|-\rangle _{C}.
\label{arb state}
\end{equation}
She makes a filtering measurement on the two particle state composed of her
part of the entangled state and the unknown state. This measurement yields
one of the four orthogonal Bell states for the pair AC 
\begin{eqnarray}
|\Psi ^{(\pm )}\rangle _{AC} &=&\sqrt{\frac{1}{2}}\left( |+\rangle
_{A}|-\rangle _{C}\pm |-\rangle _{A}|+\rangle _{C}\right)  \nonumber \\
|\Phi ^{(\pm )}\rangle _{AC} &=&\sqrt{\frac{1}{2}}\left( |+\rangle
_{A}|+\rangle _{C}\pm |-\rangle _{A}|-\rangle _{C}\right)
\label{Bell states}
\end{eqnarray}
After her measurement is completed, the particle in Bob's hands is in a
definite state depending on which result Alice obtained. Therefore, if Alice
knows the state she is given, she can view this procedure as the preparation
of one of four definite state in Bob's laboratory. Of course, Alice cannot
predict which of the four states will be produced before she makes her
measurement. That Bob ends up with this state is perfectly understandable
mathematically; however, the interpretation of what has happened is
controversial since it takes us into questions of the epistemology of
quantum mechanics. The particle in Bob's laboratory goes from having no
state to a definite state with only local measurements being performed by
Alice. This is a stark example the non-local nature of quantum theory.

In this paper, I want to discuss the mundane issues of experiments like this
and ask if a part of an entangled state is measured by a detector with
finite time resolution, what is the state of the ``undisturbed'' part system.

\section{ States of a system}

We must be more precise in defining what it means for a system to be in a
definite state \cite{Peres}. We wish to argue that a single preparation
procedure produces a definite state, but to do so the procedure must be
tested a number of times. For a pure state, the testing procedure means that
there are measurements, which can be idealized as projection measurements,
such that 
\begin{equation}
P|\phi \rangle =|\phi \rangle  \label{Projective measurement}
\end{equation}
for each realization of the procedure that produces the state $|\phi \rangle
.$ For example, if we wish to prepare an electron with its spin up along
some axis, then a Stern-Gerlach measurement along that axis is a physical
realization of $P.$ If the prepared state is $|\phi ^{\prime }\rangle \neq
|\phi \rangle ,$ then $(1-P)|\phi ^{\prime }\rangle \neq 0.$

For a mixed state, $\rho $, $\rho ^{2}\neq \rho ,$ the situation is even
more complicated. It is not sufficient to have a filtering measurement,
idealized as a complete set of orthogonal projections $\{P_{j}\},$ $%
\sum_{j}P_{j}=1,$ $P_{i}P_{j}=\delta _{ij}P_{j.}$ If we repeat the
preparation many times, the result $j$ occurs with frequency approaching $%
p_{j}=tr\rho P_{j}$, but there are an infinite number of density matrices $%
\rho $ with diagonal entries $\{$ $p_{j}\}$ $.$ Therefore, the prescription
for checking whether the prepared state is $\rho $ requires a set of
measurements that determine the off-diagonal elements of $\rho .$ The
important point is that in principle there is a method of testing a given
procedure to determine if each time it is preformed it produces the state $%
\rho $\cite{Peres}$.$ Having done this, we are allowed to argue that a
single such procedure will produce the state $\rho .$ Of course, in
practice, we are much less rigorous, relying on theory and a few
measurements to argue that a given state is prepared.

The generalization from projective measurements to positive operator valued
measurements (POVM) \cite{Peres}, \cite{Helstrom} is not difficult. In fact,
the measurements that are discussed below are more closely related to
general POVM's than to projective measurements.

\section{Preparation of a one particle state from a two particle entangled
state}

\subsection{Idealized case}

For the idealized case, we assume that idealized projection measurements can
be made instantaneously. Let $H_{1}\ $and $H_{2}$ be Hilbert spaces and
consider the space defined by their direct product. Suppose we have a
normalized bipartite state 
\begin{equation}
|\Psi \rangle =\sum_{a}c_{a}|\phi _{a}\rangle _{1}|\psi _{a}\rangle _{2},
\label{Ideal case state vector}
\end{equation}
where $\{|\phi _{a}\rangle _{1},a=1,2,\cdots \}$ is an orthonormal basis of $%
H_{1}$ and $\{|\psi _{a}\rangle _{2},a=1,2,\cdots \}$ is an orthonormal
basis of $H_{2}$. If the outcome of an idealized filtering measurement of
the complete set of projection operators $\{|\phi _{a}\rangle _{11}\langle
\phi _{a}|\}$ gives the result $a=r$ the state of particle $2$ is
instantaneously projected into the state $|\psi _{r}\rangle _{2}.$ This is
sometimes referred to as the collapse of the wave function.. The acausal
behavior of quantum theory is inherent in the fact that we cannot predict,
in principle, which $r$ the measurement of $1$ will yield. The non-local
nature of quantum mechanics is displayed by particle $2$ going from not
being in a definite state to being in a definite state even if it is far
away from particle $1.$ In a realistic theory, such as Bohm's theory \cite
{Bohm}, for each realization of the experiment, particles $1$ and $2$ have
definite trajectories determined in part by a non-local quantum potential
acting between the particles. When we determine the trajectory on which
particle $1$ lies, the trajectory of particle $2$ will be altered because
the non-local potential acting on it changes.

It is well-known that there is no superluminal signal in this case, nothing
has been transferred by the measurement of particle $1$ to the neighborhood
of particle $2$ until a signal from the output of measuring apparatus $1$
reaches $2.$ In other words, as soon as measurement $1$ is completed the
detector at $1$ has acquired $-\sum_{a}|c_{a}|^{2}\log _{2}|c_{a}|^{2}$ bits
of information. The same amount of information can be acquired by detector
in the location of $2$ by either measuring the state of particle $2$ or
receiving a signal from detector $1$ containing the result of the
measurement.

Now consider a less ideal case in which the measurement on $1$ is a POVM, $E$
. After the measurement, the state of $2$ is given by the density matrix 
\begin{equation}
\rho _{2}=\frac{1}{N}\sum_{aa^{\prime }}|\psi _{a}\rangle _{2}\left(
_{1}\langle \phi _{a}|E|\phi _{a^{\prime }}\rangle _{1}c_{a}c_{a^{\prime
}}^{*}\right) _{2}\langle \psi _{a^{\prime }}|,  \label{idealizedPOVM}
\end{equation}
where 
\[
N=\sum_{a}\text{ }_{1}\langle \phi _{a}|E|\phi _{a}\rangle _{1}|c_{a}|^{2}. 
\]
In general, this is a mixed state. Only in the special case that $%
_{1}\langle \phi _{a}|E|\phi _{a^{\prime }}\rangle _{1}$ factors into $
f_{a}f_{a^{\prime }}^{*}$ is $\rho _{2}$ a pure state. This is shown in
appendix1.

It is obvious that $\rho _{2}|\psi _{a}\rangle _{2}=0$ for any $a$ such that 
$c_{a}=0.$ This limits the state that can be prepared by measuring particle $%
1.$ This is important in the generalization of teleportation. In order for
it to be possible to teleport a state, that state must be present in the
entangle state shared by Alice and Bob.

\subsection{Finite time measurements}

\subsubsection{Detector operators}

The discussion that follows will be given in terms of the Heisenberg
picture, but it is not difficult to convert to a Schr\"{o}dinger picture. We
shall treat the particles as photons, although the conversion to any other
type of particle is not difficult. We start by specifying the measuring
devices. According to Glauber \cite{Glauber}, the detector operator for a
photon linearly polarized along the ${\bf e}$ direction is, the positive
frequency electric field operator ${\bf E=}E{\bf e}$ defined by 
\begin{equation}
E=\sum_{q}p(q,{\bf e})e^{-iq\left( t-x\right) }a(q,{\bf e})
\label{Detector operator}
\end{equation}
where $a(q,{\bf e})$ is the destruction operator for a photon linearly
polarized in the ${\bf e}$ direction with frequency $q>0$. The time is
measured in distance units so that the speed of light is one. We shall
ignore the components of momentum in the plane of the detector surface and
take $x$ to be the coordinate normal to the detector surface. We idealize to
a point detector located at $x$ that registers a count at time $t.$

To further understand this expression, let a photon in the state 
\[
|\phi \rangle =\sum_{k}f(k)a^{\dag }(k,{\bf e}^{\prime })|0\rangle 
\]
impinge on the detector. Then, using the commutation relations 
\begin{equation}
\left[ a(k,{\bf e}),a^{\dag }(k^{\prime },{\bf e}^{\prime })\right] =\delta
_{kk^{\prime }}d({\bf e},{\bf e}^{\prime }),  \label{commutation relations}
\end{equation}
where $d$ is the scalar product , 
\begin{equation}
d({\bf e},{\bf e}^{\prime })={\bf e}\cdot {\bf e}^{\prime },
\label{d function}
\end{equation}
we get 
\[
\langle 0|E|\phi \rangle =\sum_{k}f(k)p(k,{\bf e})e^{-ik(t-x)}d({\bf e},{\bf %
e}^{\prime }). 
\]
The amplitude for detection at time $t$ is in the form of a wave packet
evaluated at $x$ the location of the detector.

The detector records a quantity proportional to the intensity or,
equivalently, the counting rate, 
\begin{eqnarray}
I &=&\frac{1}{T_{m}}\int_{T-T_{m}/2}^{T+T_{m}/2}d\tau |\langle 0|E|\phi
\rangle |^{2} \\
&=&\sum_{kk^{\prime }}f(k^{\prime })^{*}p(k^{\prime },{\bf e})^{*}f(k)p(k,%
{\bf e})e^{i(k^{\prime }-k)T}\text{sinc}\left( (k-k^{\prime })\frac{T_{m}}{2}
\right) d^{2}({\bf e},{\bf e}^{\prime }),  \label{Intensity}
\end{eqnarray}
where from here on we introduce the retarded time $\tau =t-x.$ The outcome
of the measurement depends on $f,p$ and $T_{m}.$ The duration of the
measurement $T_{m}$ determines the degree to which off-diagonal matrix
elements of the state are detected. The function $p$ determines spectral
region to which the detector is sensitive.

First, suppose that the spectral amplitude $f(k)$ is peaked at $k=K$ and has
a width of $\Delta k<<K.$ Also let the width of $p$ be large compared to
that of $f$, so that $p$ is approximately constant over the range $\Delta k.$
Under these assumptions, $I$ depends on the parameter $\theta =$ $\Delta
kT_{m}=T_{m}/T_{k},$ where $T_{k}=1/\Delta k$ is the width of the wave
packet. From fig. 1 it can be seen that if $\theta \ll 1,$ the sinc function
can be replaced by $1$ over the range of summation and (\ref{Intensity})
becomes 
\[
I=|p(K,{\bf e})\sum_{\kappa }f(K+\kappa )e^{-i\kappa T}d({\bf e},{\bf e}
^{\prime })|^{2}, 
\]
where 
\begin{equation}
k=K+\kappa .  \label{expansion of k}
\end{equation}
This means that we can resolve the envelope of the wave packet by moving the
detector with respect to the source. This is illustrated in the space-time
diagram in fig. 2a.

If $\theta >>1,$ then the sinc function restricts the integration region to $%
k\approx k^{\prime }$ and (\ref{Intensity}) becomes 
\[
I=\pi |p(K,{\bf e})|^{2}\sum_{k}|f(k)|^{2}d^{2}({\bf e},{\bf e}^{\prime }). 
\]
This is the usual case for single photon detectors. This is illustrated in
fig. 2b.

Let us reverse the roles of $p$ and $f,$ so the detector has a narrow
bandwidth compared to the state. Assume that $p$ is peaked at $K_{p}$ with
width $\Delta k_{p}<<K_{p},$ such that $f$ is approximately constant over
the range $\Delta k,$ then we get a similar result with $p$ and $f$
interchanged. In this case, the quantity $I$ is determined by the detection
function $p$ and the parameter $\theta _{p}=\Delta k_{p}T_{m}.$ If $\theta
_{p}>>1,$ then 
\[
I=\pi |f(K,{\bf e})|^{2}\sum_{k}|p(k)|^{2}d^{2}({\bf e},{\bf e}^{\prime }) 
\]
and the measured intensity depends on a single mode of the particle wave
packet, fig. 2c. This case corresponds to placing a narrow filter in front
of the detector and is often used in practice.

\subsubsection{Two particle entangled states}

Now suppose that a two photon entangled state is generated with one photon
moving to the right and the other to the left, 
\begin{equation}
|\Psi \rangle =\sum_{kK}f(k,K)\left( \xi _{+}|k{\bf e}_{+}\rangle _{R}|K{\bf %
e}_{-}\rangle _{L}+\xi _{-}|k{\bf e}_{-}\rangle _{R}|K{\bf e}_{+}\rangle
_{L}\right) .  \label{two particle entangled}
\end{equation}
The linear polarization states are defined with respect to the orthogonal
directions ${\bf e}_{+}$ and ${\bf e}_{-}$. The factors $\xi _{\pm }$ are
taken to be phase factors, $|\xi _{\pm }|=1$ so that $|\Psi \rangle $ is a
superposition of plane wave Bell states like those defined in (\ref{Bell
states}). We shall assume that $f(k,K),$ the spectral amplitude, is peaked
around $k_{0}$ and $K_{0}$ with widths $\Delta k<<k_{0}$ and $\Delta
K<<K_{0}.$ This ensures that the single photon state for R, which has the
spectral function $\sum_{K}|f(k,K)|^{2},$ is a quasimonochromatic wave
packet, and ,similarly, the single photon state for L is quasimonochromatic.

Let us now detect the right-moving photon, R, at time $t_{1}$ and the
left-moving photon, L, at time $t_{2}$. The correlation function for this is
given by 
\begin{equation}
C_{12}=\langle \Psi |E_{1}^{\dag }E_{2}^{\dag }E_{2}E_{1}|\Psi \rangle ,
\label{Correlation function}
\end{equation}
with the detector operators given by eq. (\ref{Detector operator}).

It is unrealistic to assume that the measurements occur instantaneously, so
we compute 
\begin{equation}
\overline{C}_{12}=\int dt_{2}\int dt_{1}C_{12}S(t_{2},t_{1}),
\label{integrated correlation}
\end{equation}
where $S$ is one when the detectors are on and vanishes when they are off.
In the usual coincident counting experiments, $S\ $is a function of $%
t_{2}-t_{1}$ that is nonvanishing over some, usually small, time interval.
In this paper we are interested in the case $t_{2}>>t_{1},$ so that we can
ascribe meaning to the state of L in the time between the two measurements.

In the example we are considering, the correlation function becomes 
\begin{equation}
C_{12}=|A_{12}|^{2},  \label{amp1}
\end{equation}
where the two particle amplitude is 
\begin{eqnarray}
A_{12} &=&\langle 0|E_{2}E_{1}|\Psi \rangle  \nonumber \\
&=&\sum_{K}g_{1}(K)\langle 0|E_{2}|K\text{ }\widetilde{{\bf e}}_{1}\rangle
_{L},  \label{Amplitude}
\end{eqnarray}
with 
\begin{equation}
g_{1}(K)=\sum_{k}p_{R}(k,{\bf e}_{1})e^{-ik\tau _{1}}f(k,K),
\label{g-function}
\end{equation}
and the polarization state is 
\begin{equation}
|\widetilde{{\bf e}}_{1}\rangle _{L}=\sum_{\sigma =\pm }|{\bf e}_{\sigma
}\rangle _{L}\xi _{-\sigma }d({\bf e}_{1},{\bf e}_{-\sigma }).
\label{spin state}
\end{equation}
If $\xi _{+}=-\xi _{-},$ the state $|\widetilde{{\bf e}}_{1}\rangle _{L}$ is
orthogonal to $|{\bf e}_{1}\rangle _{L}.$ After the measurement of R is
completed, the photon L has a definite polarization state.

The first detector is a trigger which registers in a time interval $(T_{1}-%
\frac{T_{m}}{2},T_{1}+\frac{T_{m}}{2}).$ After detector one fires, the
correlation function reduces to a single particle function 
\begin{equation}
C_{1}=N\sum_{K,K^{\prime }}\langle 0|E_{2}|K\text{ }\widetilde{{\bf e}}%
_{1}\rangle _{L}\langle 0|E_{2}|K^{\prime }\text{ }\widetilde{{\bf e}}
_{1}\rangle _{L}^{*}\rho _{L}(K;K^{\prime })  \label{C1}
\end{equation}
where 
\begin{eqnarray}
\rho _{L}(K;K^{\prime }) &=&\frac{1}{N}\sum_{kk^{\prime }}p_{R}(k,{\bf e}
_{1})p_{R}(k^{\prime },{\bf e}_{1})^{*}f(k,K)f(k^{\prime },K^{\prime
})^{*}e^{-ikT_{1}}\times  \nonumber \\
&&e^{ik^{\prime }T_{1}}T_{m}\text{sinc}(k-k^{\prime })\frac{T_{m}}{2}.
\label{density matrix element}
\end{eqnarray}
$\rho _{L}(K;K^{\prime })$ is a matrix element of the one particle density
matrix for L. The normalization $N$ is defined so that $tr\rho _{L}=1.$
Finally, we have 
\begin{equation}
C_{1}=N\text{ }tr\left( \rho _{L}E_{2}^{\dag }E_{2}\right)
\label{C1 density matrix form}
\end{equation}
where 
\begin{equation}
\rho _{L}=\sum_{KK^{\prime }}|K,\widetilde{{\bf e}}_{1}\rangle _{L\text{ }
}\rho _{L}(K;K^{\prime })_{L}\langle K^{\prime },\widetilde{{\bf e}}_{1}|.
\label{density matrix 1}
\end{equation}

We now investigate under what circumstances this density matrix represents a
pure state. To do this we exploit the assumption that $f$ satisfies the
condition that its width $\Delta k<<k_{0}$ and use eq.(\ref{expansion of k})
. The sinc function is small unless $|\kappa -\kappa ^{\prime }|<2\pi
/T_{m}. $ We introduce, as we did above, $T_{k}=1/\Delta k,$ the single
particle coherence time of the wave packet of particle R. The critical
parameter for the following discussion is $\theta =$ $T_{m}/T_{k}.$ As we
did in section B1 above, it is simplest to consider the two extreme cases of
long triggering times $\theta \gg 1$ and short triggering times $\theta \ll
1.$ We shall see that in the first case L is , in general, in a mixed state,
while in the second case, L is always in a pure state.

\paragraph{Long triggering times}

For long triggering times the sinc function is non-negligible when $|\kappa
-\kappa ^{\prime }|< \Delta k/\theta<<\Delta k.$ In this case, as
illustrated in fig. \ref{fig1}, we may set $\kappa \approx \kappa ^{\prime }$
in $f$ and obtain 
\begin{equation}
\rho _{L}(K;K^{\prime })=\sqrt{\frac{1}{N^{\prime }}}\sum_{k}|p_{R}(k,{\bf e}%
_{1})|^{2}f(k,K)f(k,K^{\prime })^{*}.  \label{rho long}
\end{equation}
In general, L is in a mixed state. We will discuss this further below.

\paragraph{Short triggering times}

In this case the width of the sinc function, $2\pi /T_{m}=2\pi /\left(
T_{k}\theta \right) $ $\gg $ $\Delta k,$ and the sinc function may be set
equal to $1$ over the entire range of the summation over $\kappa $ and $%
\kappa ^{\prime }.$ Consequently, 
\begin{equation}
\rho _{L}(K;K^{\prime })=\chi (K)\chi (K^{\prime })^{*}  \label{rho short}
\end{equation}
\begin{equation}
\chi (K)=\sqrt{\frac{1}{N}}\sum_{k}p_{R}(k,{\bf e}_{1})f(k,K)e^{-ik\tau
_{1}}.  \label{state short}
\end{equation}
So that $\rho _{L}(K;K^{\prime })$ factors. In this case, eq.(\ref{C1})
becomes 
\begin{equation}
C_{1}=N|\langle 0|E_{L}(x_{2},t_{2})|\chi ,\widetilde{{\bf e}}_{1}\rangle
_{L}|^{2}.  \label{C1 short}
\end{equation}
We interpret this as staying that upon completion of the measurement on R, L
is put in the pure state 
\begin{equation}
|\chi ,\widetilde{{\bf e}}_{1}\rangle _{L}=\sum_{K}\chi (K)|K,\widetilde{%
{\bf e}}_{1}\rangle _{L}.  \label{state vector short}
\end{equation}

\paragraph{The properties of the state of the left moving photon}

The exact nature of the state of the particle moving to the left depends
upon the initial entangled state and the measurement made on the right. For
long triggering time, case (a) above, the explicit time of the first
measurement has disappeared from the calculation. Of course, it is still
present in that any measurement on L must be made after the first
measurement is completed. This information is hidden by the fact that we did
not include the corrections due to the width of sinc function but treated it
as though it were a Dirac delta function.

Suppose that the measurement on the right was a filtering measurement in $k$
so that $p_{R}(k,{\bf e}_{1})$ is narrowly peaked at $k_{0}.$ This reduces
to the short triggering time, case (b), and (\ref{rho short}) holds with 
\begin{equation}
\rho _{L}(K;K^{\prime })=\frac{1}{N^{^{\prime \prime }}}%
f(k_{0},K)f(k_{0},K^{\prime })^{*}  \label{density matrix filtered long}
\end{equation}
so eq. (\ref{state vector short}) becomes 
\begin{equation}
|\chi ,\widetilde{{\bf e}}_{1}\rangle _{L}=\sqrt{\frac{1}{N^{\prime \prime }}%
}\sum_{K}f(k_{0},K)|K,\widetilde{{\bf e}}_{1}\rangle _{L}.
\label{filtered state vector}
\end{equation}
This is the case in which detector 1 has a narrow filter in front of it so
that it projects a plane wave state of R. In general, the state of L is not
a plane wave state.

A particularly interesting example of the entangled two particle state (\ref
{two particle entangled}) is the one contemplated by Einstein, Rosen and
Podolsky \cite{EPR} (EPR), 
\begin{equation}
f(k,K)=v(K)\delta (k+K-k_{p})  \label{downconversion}
\end{equation}
where $k_{p}=k_{0}+K_{0}.$ Our assumptions imply that $v(K)$ is peaked
around $K_{0}.$ Such a two-particle entangled state can approximately be
realized for photons using type-II spontaneous parametric down-conversion
for which $k_{p}$ is the pump frequency.

In case (a ), 
\begin{equation}
\rho _{L}(K;K^{\prime })=\frac{1}{N}|p_{R}(k_{p}-K,{\bf e}%
_{1})|^{2}|v(K)|^{2}\delta (K-K^{\prime }),
\label{density matrix down conversion}
\end{equation}
so that $\rho _{L}$ is diagonal in the basis of $|K$ $\widetilde{{\bf e}}%
_{1}\rangle $ states. The state for case (b) becomes a plane wave state.

If the filter function $p_{R}$ is narrowly peaked at $k_{0},$ so that we
have case (b) again. The state of the L photon has a spectrum determined by $%
f(k_{0},K)$ which in turn is fixed by the original two-photon state. The
result of the measurement selects the state of particle L.

A contentious issue in the interpretation of quantum mechanics is whether
the uncertainty principle reflects a fundamental limitation on how well
conjugate variables can be determined because of the basic quantum nature of
measurement, as Heisenberg believed \cite{Heisenberg}. This position was
criticized by Popper \cite{Popper} who argued that the uncertainty principle
was a statistical statement and did not imply that it was meaningless for a
particle to simultaneously possess definite values of conjugate variables as
they do classically. It is clear that the measurement of the uncertainty in
the left moving particle is unchanged by the measurement of R if the
uncertainty is computed based on all the photons moving to the left (that
is, independently of whether R registers a count or not). On the other hand,
if the uncertainty is measured only for those L photons whose partners are
detected on the right, then the uncertainty is different. The measurement
changes the uncertainty because it selects out a subset of the particles
moving to the left. There is no action-at-a-distance in the sense of a force
changing the uncertainty. The possible outcomes of individual experiments
and the statistics of sets of experiments come from the original entangled
state through $f.$ The non-local action occurs for each individual
experiment, so that after the detection on the right, the photon moving to
the left has gone from not having a definite state to having a definite
state. The uncertainty changes because the experiment dictates that we
compute it with a conditional probability.. Only those states of the
particle $L$ are considered which are associated with the triggering of the
right detector and this conditioning depends on the nature of the detector
through $p_{R}$ and $T_{M}$ \cite{comment}.

The fact that measurements do not necessarily induce uncontrolled
uncertainty in the sense of Heisenberg is by now well-known from the
discussion of the quantum erasure \cite{Scully}.

\section{Measurement of three particle states}

We shall consider a three particle state that is the product of an entangled
two particle state and an independent one particle state. This type of state
is discussed in the original teleportation paper \cite{teleportation}. In
that case, the measurement is performed on two of the particles and a single
particle state is prepared. Alternatively, by measuring one of the particles
in the entangled state, one can prepare a state of the two remaining
particles. A problem related to this case has been discussed by Horne \cite
{Horne} in connection with measuring one particle in a four particle state
to produce a GHZ state \cite{Zeilinger}.

The state we will consider is 
\begin{equation}
|\Psi \rangle =|\Psi \rangle _{ab}|\Phi \rangle
_{c}=\sum_{k_{a}k_{b}}f(k_{a},k_{b})\left( |k_{a}{\bf e}_{+}\rangle
_{a}|k_{b}{\bf e}_{-}\rangle _{b}+|k_{a}{\bf e}_{-}\rangle _{a}|k_{b}{\bf e}
_{+}\rangle _{b}\right) \sum_{k_{c}}g(k_{c})|\phi ;k_{c}\rangle _{c}
\label{three particle entangled}
\end{equation}
where 
\begin{equation}
|\phi ;k_{c}\rangle _{c}=\alpha _{+}|k_{c}{\bf e}_{+}\rangle _{c}+\alpha
_{-}|k_{c}{\bf e}_{-}\rangle _{c}  \label{unknown state}
\end{equation}
is a normalized plane wave state. The two particle entangled state is not
the most general such state, but is rather a superposition of the plane wave
entangled states similar to the one Bohm used in his discussion of the EPR
experiment \cite{Bohmtext}.

\subsection{Measurement of the Bell states}

For quantum teleportation it is necessary to perform a measurement that
projects the state of the particles $a$ and $c$ onto the Bell states (in
this section lower case $a$ refers to Alice and lower case $b,$ to Bob,
upper case $B,$ to Bell). To do this it is necessary to define a Bell state
detector operator. The four Bell states are defined by 
\begin{equation}
|B;k_{1},k_{2}\rangle _{ab}=\sqrt{\frac{1}{2}}\sum_{\sigma =\pm }\text{ }%
\sum_{\mu =\pm }\zeta _{\sigma \mu }^{(B)}|k_{1}{\bf e}_{\sigma }\rangle
_{a}|k_{2}{\bf e}_{\mu }\rangle _{b}  \label{Bell states plane wave}
\end{equation}
where the non-zero elements of the $\zeta $ are $\zeta _{++}^{(1)}=\zeta
_{--}^{(1)}=1,$ $\zeta _{++}^{(2)}=-\zeta _{--}^{(2)}=1,$ $\zeta
_{+-}^{(3)}=\zeta _{-+}^{(3)}=1,$ and $\zeta _{+-}^{(4)}=-\zeta
_{-+}^{(4)}=1,$ and ${\bf e}_{+}$ and ${\bf e}_{-}$ are orthogonal
polarization vectors . The Bell state detector operator is defined as

\begin{equation}
E^{(B)}=\sum_{k_{1}k_{2}}p^{(B)}(k_{1},k_{2})e^{-i\left( k_{1}+k_{2}\right)
\tau _{B}}\sum_{\sigma \mu }\zeta _{\sigma \mu }^{(B)}a(k_{1},{\bf e}
_{\sigma })a(k_{2},{\bf e}_{\mu }).  \label{Bell detector}
\end{equation}
The retarded time $\tau _{B}=t_{B}-x_{B,}$ where $x_{B}$ is the coordinate
normal to the detector and $t_{B}$ is the time the detector registers the
pair. We have chosen the form of the detector based on a model in which
up-conversion is used to detect the Bell states.

Following \cite{teleportation}, we rewrite (\ref{three particle entangled})
as 
\begin{equation}
|\Psi \rangle
=\sum_{k_{a}k_{b}k_{c}}f(k_{a},k_{b})g(k_{c})\sum_{B}|B;k_{a},k_{c}\rangle
_{ac}|\phi ^{(B)};k_{b}\rangle _{b},  \label{Teleportation state}
\end{equation}
where $|\phi ^{(B)};k_{b}\rangle _{b}$ is the plane wave state associate
with $B$ in Bob's laboratory. It is related to (\ref{unknown state}) by a
spin transformation $\Lambda _{B}\phi =\phi ^{(B)},$ \cite{teleportation}.

Now suppose we measure the three particle correlation 
\begin{equation}
C_{B}=\langle \Psi |E^{(B)\dag }E_{b}^{\dag }E_{b}E^{(B)}|\Psi \rangle
=|A_{bB}|,  \label{correlation for bell}
\end{equation}
where the amplitude $A_{bB}$ is given by 
\begin{eqnarray}
A_{bB} &=&\langle 0|E_{b}E^{(B)}|\Psi \rangle  \nonumber \\
&=&\sum_{k_{a}k_{b}k_{c}}U_{B}(k_{a},k_{b},k_{c})e^{-i\left(
k_{a}+k_{c}\right) \tau _{B}}\langle 0|E_{b}|\phi ^{(B)};k_{b}\rangle _{b},
\label{Bell amplitude}
\end{eqnarray}
and 
\begin{equation}
U_{B}(k_{a},k_{b},k_{c})=f(k_{a},k_{b})g(k_{c})p^{(B)}(k_{a},k_{c}).
\label{U function}
\end{equation}

The procedure is now the same as above, we integrate $C_{B}$ over the
detection time $t_{B}$ of the Bell state detector in Alice's laboratory,
call the result $\overline{C}_{B}.$ The integration gives a sinc function
that depends on the energies of the particles in Alice's laboratory, 
\begin{equation}
S_{ac}=\text{sinc}\left( k_{a}+k_{c}-k_{a}^{\prime }-k_{c}^{\prime }\right) 
\frac{T_{m}}{2}.  \label{sinc1}
\end{equation}
Finally, we express $\overline{C}_{B}$ in terms of a density matrix for the
particle in Bob's laboratory, 
\begin{equation}
\overline{C_{B}}=N\sum_{k_{b}k_{b}^{\prime }}\langle 0|E_{b}|\phi
^{(B)};k_{b}\rangle _{b}\langle 0|E_{b}|\phi ^{(B)};k_{b}^{\prime }\rangle
_{b}^{*}\rho _{B}(k_{b},k_{b}^{\prime }),  \label{CB}
\end{equation}
with 
\begin{equation}
\rho _{B}(k_{b},k_{b}^{\prime })=\frac{1}{N}\sum_{k_{a}k_{b}k_{a}^{\prime
}k_{b}^{\prime }}S_{ac}e^{-i\left( k_{a}+k_{c}\right)
T_{B}}U_{B}(k_{a},k_{b},k_{c})e^{i\left( k_{a}^{\prime }+k_{c}^{\prime
}\right) \tau _{B}}U_{B}(k_{a}^{\prime },k_{b}^{\prime },k_{c}^{\prime
})^{*},  \label{Bobs density matrix}
\end{equation}
where, as usual, $N$ is a normalization constant.

We are interested in accurate teleportation, so we want Bob's state to be a
pure state. To this end, we require that $S_{ac}$ in eq.(\ref{sinc1}) to be
approximately equal to one over the range of integration. This entails that 
\begin{equation}
\left( \Delta k_{a}+\Delta k_{c}\right) T_{m}<<2\pi ,
\label{width condition}
\end{equation}
where $\Delta k_{a}$ is the width of $f(k_{a},k_{b})$ in the first variable
and $\Delta k_{c}$ is the width of $g(k_{c}).$ If (\ref{width condition}) is
satisfied, when the outcome of Alice's Bell state measurement is $B,$ (\ref
{CB}) becomes 
\[
\overline{C_{B}}=\text{ }_{b}\langle \chi ^{(B)}|E_{b}^{\dag }E_{b}|\chi
^{(B)}\rangle _{b}. 
\]
The state produced in the Hilbert space of Bob's particle is 
\begin{equation}
|\chi ^{(B)}\rangle _{b}=\sqrt{\frac{1}{N}}\sum_{k_{b}}\left[
\sum_{k_{a}k_{b}}U(k_{a},k_{b},k_{c})e^{i\left( k_{a}+k_{c}\right)
T_{B}}\right] |\phi ^{(B)};k_{b}\rangle _{b}.  \label{Bobs state}
\end{equation}
This is a pure state but, in general, does not have the same spectral
properties of the original function, that is, it is not equal to 
\begin{equation}
|\Phi _{B}\rangle _{b}=\sum_{k_{b}}g(k_{b})|\Lambda _{B}\phi ;k_{b}\rangle
_{b}.  \label{Bob's desired state}
\end{equation}
For accurate teleportation we require that these be equal up to a phase.
This leads to the condition 
\begin{equation}
\sum_{k_{c}}\left( \sum_{k_{a}}f(k_{a},k_{b})p^{(B)}(k_{a},k_{c})e^{i\left(
k_{a}+k_{c}\right) T_{B}}\right) g(k_{c})=\lambda g(k_{b}).
\label{eigenvalue eq}
\end{equation}
This equation requires that $g$ be an eigenvector of the operator in
brackets. This operator depends on the input entangled state, $f$ , and the
nature of the Bell state detector, $p^{(B)}.$ Note that the functional
dependence on $k_{b}$ appears in $f$ . This indicates that $f$ limits the
class of functions that can be teleported.

If there are approximations such that the operator in (\ref{eigenvalue eq})
is a constant times the identity matrix, any $g$ consistent with (\ref{width
condition}) and these approximations can be accurately teleported. As an
example of such a case, let $f$ to be given by eq.(\ref{downconversion}). In
this case (\ref{eigenvalue eq}) becomes 
\[
\sum_{k_{c}}\left( \nu (k_{b})p^{(B)}(k_{p}-k_{b},k_{c})e^{i\left(
k_{p}-k_{b}+k_{c}\right) T_{B}}\right) g(k_{c})=\lambda g(k_{b}). 
\]
In addition, if 
\begin{equation}
p^{(B)}(k_{a},k_{c})=p_{0}^{(B)}\delta (k_{a}+k_{c}-k_{0}),
\label{factorization}
\end{equation}
then eq. (\ref{eigenvalue eq}) becomes 
\[
p^{(B)}\nu (k_{b})e^{ik_{0}T_{B}}g(k_{0}-k_{p}+k_{b})=\lambda g(k_{b}). 
\]
Finally, take $k_{0}=k_{p}$. We now can satisfy eq. (\ref{eigenvalue eq})
for the class of $g(k_{b})$ such that $v(k_{b})$ is approximately constant
over the domain of $k_{b}$ where $g(k_{b})$ is non-zero. The condition $%
k_{0}=k_{p}$ is a requirement on the detector function. For an up-conversion
model of the Bell state detector, this means that the up-converted photon
has the same energy as the pump photon that produced the original entangled
state $|\Psi \rangle _{ab}$ in eq.~(\ref{three particle entangled}). In
practice the approximation made here restricts the class of states that can
be accurately teleported to quasimonochromatic states. On the other hand,
for given $f$ and $p^{(B)}$ the eigenvalue equation (\ref{eigenvalue eq})
may have a richer set of solutions that permit accurate teleportation.

There is an assumption in our discussion that requires further
consideration. We have assumed that the spectral functions of the entangled
state and the single particle states are the same for each realization of
the experiment. Usually this will not be true \cite{Walther}. To see how
this effects the outcome we consider a simple case. Suppose that for each
experimental realization, $f$ is the same and $g$ has a phase factor that
varies from experiment to experiment. This might be due to the generation of
the single particle state at different optical distances from the entangled
pair. For the $jth$ experiment suppose that 
\begin{equation}
g^{(j)}(k)=e^{i\Theta _{j}(k)}g(k).  \label{phase effect}
\end{equation}
Now we must average over $j$ to compute the density matrix. This gives 
\[
\langle \rho _{B}(k_{b},k_{b}^{\prime })\rangle =\sum_{j}\int D\Theta
_{j}(k)p(\Theta _{j}(k))\rho _{B}^{(j)}(k_{b},k_{b}^{\prime }), 
\]
where $p(\Theta (k))$ is the probability distribution function for $\Theta
(k),$ $\rho _{B}^{(j)}$ is given by eq.'s (\ref{Bobs density matrix}) and (%
\ref{U function}) with $g$ replaced by $g^{(j)}.$ If $\Theta _{j}$ is
independent of $k,$ then $\langle \rho _{B}\rangle =\rho _{B}.$ On the other
hand, suppose $\Theta _{j}(k)$ is random and that $\langle e^{i\left( \Theta
_{j}(k)-\Theta _{j}(k^{\prime })\right) }\rangle =\delta _{kk^{\prime }}.$
In this case, Bob's state is determined by the condition 
\[
\langle \overline{C_{B}}\rangle =NtrE_{b}^{\dag }E_{b}\langle \rho
^{(B)}\rangle 
\]
giving 
\[
\langle \rho ^{(B)}\rangle =\frac{1}{N}\sum_{k_{b}k_{b}^{\prime }}|\phi
^{(B)};k_{b}\rangle _{b}\left( \sum_{k_{a}k_{a}^{\prime }k_{c}}e^{-i\left(
k_{a}-k_{a}^{\prime }\right) T_{B}}U(k_{a},k_{b},k_{c})U(k_{a}^{\prime
},k_{b}^{\prime },k_{c})^{*}\right) \langle \phi ^{(B)};k_{b}^{\prime }|. 
\]
In general this expression will not factor, so Bob ends up with a mixed
state rather than a pure state.

To overcome this type of random phase disturbance, experimentalists usually
produce the entangled state and the state to be teleported coherently \cite
{Bouwmeester},\cite{Boschi}.

\subsection{Measurement of a single particle}

Next consider the case in which the state (\ref{three particle entangled})
is generated and a single particle is measured. A measurement of particle $a$
will not entangle $b$ and $c.$ However, suppose that we mix $b$ and $c$ by
passing the pair through a 50-50 beam splitter as illustrated in fig. 3. If $%
a$ is detected, the remaining pair will be partially entangled. The outgoing
pair will have a density matrix of the form $\rho =\rho _{12}+\rho
_{11}+\rho _{22}$ corresponding to one photon going to each detector, $\rho
_{12},$ or both photons going to the same detector, $\rho _{11}$ and $\rho
_{22}.$.

For a beam splitter with equal transmittance and reflectance, the field for
detector 1 is 
\begin{equation}
\Xi _{1}=\frac{1}{\sqrt{2}}\sum_{q}p_{1}(q,{\bf e}_{1})e^{-iq\tau
_{1}}\left( ia_{b}(q,{\bf e}_{1})+a_{c}(q,{\bf e}_{1})\right) =\frac{1}{%
\sqrt{2}}(iE_{1b}+E_{1c}).  \label{detector 1 beam splitter}
\end{equation}
The phase factor of $i$ is associated with reflection off the beam splitter.
The detector operator $E_{2}$ is similarly defined. The triple correlation
function is composed of three non-interfering terms, one in which the
particles $b$ and $c$ go to different detectors and two in which they go to
the same detector. 
\begin{equation}
C_{123}=\langle \Psi |\Xi _{1}^{\dag }\Xi _{2}^{\dag }E_{3}^{\dag }E_{3}\Xi
_{2}\Xi _{1}|\Psi \rangle =|A_{12}|^{2}+|A_{11}|^{2}+|A_{22}|^{2},
\label{correlation function three}
\end{equation}
where 
\begin{eqnarray}
A_{12} &=&\frac{1}{2}\left( \langle 0|E_{3}E_{2c}E_{1b}|\Psi \rangle
-\langle 0|E_{3}E_{2b}E_{1c}|\Psi \rangle \right)
\label{different detectors} \\
A_{11} &=&\frac{i}{2}\langle 0|E_{3}E_{1b}E_{1c}|\Psi \rangle
\label{same detector 1} \\
A_{22} &=&\frac{i}{2}\langle 0|E_{3}E_{2c}E_{2b}|\Psi \rangle .
\label{same detector 2}
\end{eqnarray}
The notation $E_{1b}$ means that the operator defined in (\ref{Detector
operator}) contains the destruction operator acting on the photon in the $b$
mode. From the point of view discussed in this paper, we must keep all these
terms in order to specify the state prepared when detector 3 registers a
count. In many discussions, the amplitude for both particles going to the
same detector is dropped on the grounds that only the coincidences of
detectors 1 and 2 are registered. It is then justified to argue that only $%
A_{12}$ is observed.

As shown in appendix 2, 
\begin{equation}
A_{12}=\frac{\sqrt{N}}{2}\langle 0|E_{1}E_{2}|\chi \rangle _{12}
\label{different detectors1}
\end{equation}
where 
\begin{equation}
|\chi \rangle _{12}=\frac{1}{\sqrt{N}}\left( |\Phi ^{\prime }\rangle
_{1}|\Phi \rangle _{2}+|\Phi \rangle _{1}|\Phi ^{\prime }\rangle _{2}\right)
.  \label{entangled two particle state}
\end{equation}
This is an entangled state composed of single particle states that are
superpositions of the plane wave states (\ref{unknown state}) and (\ref
{phiprimed}), 
\begin{eqnarray}
|\Phi \rangle &=&\sum_{k}g(k)|\phi ;k\rangle ,  \nonumber \\
|\Phi ^{\prime }\rangle &=&\sum_{k}\gamma (k)|\phi ^{\prime };k\rangle ,
\label{Omega state functions}
\end{eqnarray}
where 
\begin{equation}
\gamma (k)=\sum_{k_{3}}p(k_{3},{\bf e}_{3})e^{-ik_{3}\tau _{3}}f(k_{3},k).
\label{gamma function}
\end{equation}
The state $|\Phi \rangle $ is just the original input single particle state,
(\ref{three particle entangled}).

In an identical way 
\[
A_{11}=\frac{1}{2\sqrt{N}}\langle 0|E_{1}E_{1}|\chi \rangle _{11}, 
\]
with 
\[
|\chi \rangle _{11}=\sqrt{N}\sum_{kk^{\prime }}|\phi ;k\rangle _{1}|\phi
^{\prime };k^{\prime }\rangle _{1}g(k)\gamma (k^{\prime }). 
\]
A similar result holds for $A_{22}.$

We now repeat the calculation made in the first section of the paper. This
case is much more complicated. We get several terms 
\[
\overline{C}_{123}=C_{12}+C_{11}+C_{22} 
\]
\begin{eqnarray*}
C_{12} &=&\int_{T_{3}-\frac{T_{t}}{2}}^{T_{3}+\frac{T_{t}}{2}}d\tau
_{3}|A_{12}|^{2} \\
C_{zz} &=&\frac{1}{4}\int_{T_{3}-\frac{T_{t}}{2}}^{T_{3}+\frac{T_{t}}{2}%
}d\tau _{3}|\langle 0|E_{z}E_{z}|\chi \rangle _{zz}|^{2},\hspace{0.5in}z=1,2.
\end{eqnarray*}
More explicitly, 
\begin{eqnarray*}
C_{12} &=&W_{1}+W_{2}+2%
\mathop{\rm Re}%
W_{3}=NtrE_{1}^{\dag }E_{2}^{\dag }E_{1}E_{2}\rho _{12}, \\
C_{mm}^{(s)} &=&NtrE_{m}^{\dag }E_{m}^{\dag }E_{m}E_{z}\rho _{mm},\quad m=1,2
\end{eqnarray*}
$W_{1}$ and $W_{2}$ come from the squares of the two terms in (\ref
{different detectors}), and $W_{3}$ is the interference term between these
two amplitudes. After the integration over $\tau _{3},$ the parameter that
determine the nature of the unmeasured pair is $\theta =$ $T_{m}/T_{k}$
where $T_{k}=1/\Delta k$ , $\Delta k$ is the maximum (with respect to $k_{3}$
) width of $p(k_{3},{\bf e}_{3})f(k_{3},k)$ . If this is much less than $1,$
we are in the short trigger time limit and 
\[
\rho _{12}=|\chi \rangle _{12\text{ \thinspace }12}\langle \chi |, 
\]
The overlap between the two terms is $|\langle \Omega ^{\prime }|\Omega
\rangle |^{2}.$ In order for this to reach a maximum it is necessary that $%
|\sum_{k}g(k)\gamma ^{*}(k)|^{2}$ be a maximum or $g(k)=\lambda \gamma (k)$
for a constant $\lambda .$ This places a condition on the detector function $%
p$ for each choice of $f$ and $g.$ This case is similar to that following (%
\ref{eigenvalue eq}).

In this discussion, it has been assumed that the photons $b$ and $c$ are not
recombined after they past the first beam splitter. If this were not the
case, then the state after the beam splitter is actually a superposition of $%
|\chi \rangle _{12},|\chi \rangle _{11,}$and $|\chi \rangle _{22}.$
Furthermore, as in the case discussed in the previous section, if we let $%
g(k)$ contain a random phase, then in general the state produced after the
beam splitter will be a mixed state.

\section{Conclusion}

If we have a set of N entangled particles, the subsystems of the entangles
states are not in any definite state. The effect of measuring a subset of
the M particles is to produce a state of N-M particles. The precise nature
of this state depends on the initial entangled state and the nature of the
measurement. In particular, there is a time scale set by the initial
entangled state and the subsystem measured such that if the duration of the
measurement is long on this time scale, then the state of the N-M particles
prepared will be a mixed state. If the duration is short, then the state
prepared is a pure state. One way to ensure that the latter case holds is to
place filters in front of the measuring devices such that a definite state
of the measured subsystem is projected out by the detectors. In practice,
for photons, this is done using narrow spectral filters.

We have shown that accurate quantum teleportation can not be done for
arbitrary states and found an integral equation that the state spectral
amplitude must satisfy. This condition shows how the teleportation of states
allows local measurements in Alice's laboratory to determine a complicated
state in Bob's laboratory. In particular it shows that the spacial
information must already be present in the entangled state.

We have also seen how a measurement of one particle from an entangled pair
can lead to a partially entangled state of an independent particle and the
second particle form the pair. This is done by mixing the unmeasured pair on
a beam splitter. However, the entangled pair that is produced is not
composed of identical states in the two outputs. For this to occur, it is
again necessary that the independent state and the entangled state have
spectral amplitudes that are related.

\section{Acknowledgments}

The author wishes to express his thanks to the members of the UMBC quantum
computing seminar, especially Mark Heiligman, Keith Miller, Arthur Pittenger
and Yanhua Shih, for many stimulating discussions. This work was in part
supported by grants from the Office of Naval Research, the National Security
Agency and the Army Research Office.

\appendix

\section{Proof of the factorization of the POVM}

For a pure state $tr\rho _{2}^{2}=1.$ Using eq.( \ref{idealizedPOVM}) this
condition becomes 
\begin{eqnarray*}
1 &=&\frac{1}{N^{2}}\sum_{aa^{\prime }}|_{1}\langle \phi _{a}|E|\phi
_{a^{\prime }}\rangle _{1}|^{2}|c_{a}|^{2}|c_{a^{\prime }}|^{2} \\
&\leq &\left( \frac{1}{N}\sum_{a}\text{ }_{1}\langle \phi _{a}|E|\phi
_{a}\rangle _{_{1}}|c_{a}|^{2}\right) ^{2}=1,
\end{eqnarray*}
since 
\[
|_{1}\langle \phi _{a}|E|\phi _{a^{\prime }}\rangle _{1}|^{2}\leq \text{ }%
_{1}\langle \phi _{a}|E|\phi _{a}\rangle _{1\text{ }}\text{ }_{1}\langle
\phi _{a^{\prime }}|E|\phi _{a^{\prime }}\rangle _{1}. 
\]
The last equality is just the normalization of $\rho _{2}.$ The Schwarz
inequality becomes an equality if, and only if, 
\begin{equation}
E|\phi _{a^{\prime }}\rangle _{1}=z_{a^{\prime }a}E|\phi _{a}\rangle _{1}
\label{eqx}
\end{equation}
for a constant $z_{aa^{\prime }}.$ By taking the inner product of this
equation with first with $|\phi _{a}\rangle _{1}$ and then with $|\phi
_{a^{\prime }}\rangle _{1}$ that for each $a$ and $a^{\prime }$ 
\[
_{1}\langle \phi _{a}|E|\phi _{a^{\prime }}\rangle _{1}=e^{i\theta
_{aa^{\prime }}}\sqrt{_{1}\langle \phi _{a}|E|\phi _{a}\rangle _{1}}\sqrt{%
_{1}\langle \phi _{a^{\prime }}|E|\phi _{a^{\prime }}\rangle _{1}} 
\]
where $\theta _{aa}=0$ for all $a,$ since $E\geq 0.$

We now have 
\[
z_{aa^{\prime }}=e^{i\theta _{aa^{\prime }}}\sqrt{\frac{_{1}\langle \phi
_{a^{\prime }}|E|\phi _{a^{\prime }}\rangle _{1}}{_{1}\langle \phi
_{a}|E|\phi _{a}\rangle _{1}}} 
\]
Finally taking the inner product of $|\phi _{b}\rangle _{1}$ with \ref{eqx}
we can show that $\theta _{ba^{\prime }}=\theta _{aa^{\prime }}+\theta _{ba}$
from which it follows that $\theta _{aa^{\prime }}=\xi _{a}-\xi _{a^{\prime
}}.$

\section{Calculation of output state from beam splitter}

The Bell state given in (\ref{three particle entangled}) corresponds to $B=3$
so the first term in $A_{12}$ is given by 
\begin{equation}
\langle 0|E_{3}E_{2b}E_{1c}|\Psi \rangle
=\sum_{k_{1}k_{2}k_{3}}g(k_{1})f(k_{3},k_{2})\langle 0|E_{1}|\phi
;k_{1}\rangle _{1}\langle 0|E_{3}E_{2}|3;k_{3},k_{2}\rangle _{32},
\label{Amplitude three particles2}
\end{equation}
where the index $2$ refers to states after the beam splitter, the index $%
3=a, $ as shown in the fig. 3, and 
\[
\langle 0|E_{3}E_{2}|3;k_{3},k_{2}\rangle _{32}=p(k_{3},{\bf e}%
_{3})e^{-ik_{3}\tau _{3}}\langle 0|E_{2}|\phi ^{\prime };k_{2}\rangle , 
\]
with 
\begin{equation}
|\phi ^{\prime };k_{2}\rangle =|k_{2}{\bf e}_{-}\rangle _{2}d({\bf e}_{3},%
{\bf e}_{+})+|k_{2}{\bf e}_{+}\rangle _{2}d({\bf e}_{3},{\bf e}_{-}).
\label{phiprimed}
\end{equation}
The operators $E_{1}$ and $E_{2}$ are of the form given in (\ref{Detector
operator}). The second term in $A_{12}$ is 
\begin{equation}
\langle 0|E_{3}E_{2c}E_{1b}|\Psi \rangle
=\sum_{k_{1}k_{2}k_{3}}g(k_{2})f(k_{3},k_{1})\langle
0|E_{3}E_{1b}|3;k_{3},k_{1}\rangle _{31}\langle 0|E_{2}|\phi ;k_{2}\rangle
_{2},  \label{Amplitude three particles3}
\end{equation}
where 
\[
\langle 0|E_{3}E_{1b}|3;k_{3},k_{1}\rangle _{31}=p(k_{3},{\bf e}%
_{3})e^{-ik_{3}\tau _{3}}\langle 0|E_{1}|\phi ^{\prime };k_{1}\rangle _{1}. 
\]

Equation (\ref{different detectors1}) now follows.

\pagebreak

\begin{figure}[hb]
\caption{Illustration of the two cases $\theta >>1$ and $\theta <<1$.}
\label{fig1}
\end{figure}

\begin{figure}[htb]
\caption{The vertical lines are the world lines for the source, S, and
detector,D. The thick lines represent the limits of the signal. In c we
illustrate the effect of the filter, F, in spreading the signal.}
\label{fig2}
\end{figure}

\begin{figure}[htb]
\caption{One member, a, of the Bell state is detected at the detector D3.
The other particle, b, goes to the beamsplitter where it is mixed with the
single particle c.}
\label{fig3}
\end{figure}


\begin{references}
\bibitem{Peres}  A. Peres, {\it Quantum Theory:Concepts and Methods},
(Kluwer Academic Publishers, Dordrecht, 1993).

\bibitem{GHZ}  D. M. Greenberger, M. Horne, and A. Zeilinger, in Bell's
Theorem, {\it Quantum Theory, and Conceptions of the Universe}, M. Kaftos
ed., (Kluwer, Dordrect, 1989); D. M. Greenberger, M. Horne, A. Shimony, and
A. Zeilinger, Am. J. Phys. {\bf 50}, 1131 (1990).

\bibitem{teleportation}  C. H. Bennett, et. al., Phys. Rev. Lett. {\bf 70},
1895 (1993).

\bibitem{Helstrom}  C. W. Helstrom, {\it Quantum Detection and Estimation
Theory}, (Academic Press, New York, 1976).

\bibitem{Bohm}  D. Bohm, Phys. Rev. {\bf 85}, 166, 180 (1952).

\bibitem{Glauber}  R. J. Glauber, Phys. Rev. {\bf 130, }2529 (1963).

\bibitem{EPR}  A. Einstein, B. Podolsky, and N. Rosen, Phys. Rev. {\bf 47},
777 (1935).

\bibitem{Heisenberg}  W. Heisenberg, {\it The Physical Principles of the
Quantum Theory}, (Dover Publications, Chicago, 1930).

\bibitem{Popper}  K. Popper, {\it Quantum Theory and the Schism in Physics},
W. W. Bartley, III, ed., (Hutchinson, London, 1983). Y. H. Kim, R. Yu, and
Y. H. Shih, {\it submitted for published.}

\bibitem{comment}  The single slit experiment analyzed by Heisenberg can be
looked at in a similar fashion in which the slit makes a selection of
possible outcomes. Before reaching the slit the state of the system is a
plane wave $e^{i\omega z/c}.$ This can be expressed as $e^{i\omega
z/c}\sum_{n}f_{na}(y)$ where $f_{na}(y)$ $=1$ for $|y-na|<a/2$ and vanishes
elsewhere. The slit then projects out the term $f_{0}(y)$ with the
consequent introduction of the uncertainty in the y-component of momentum. I
believe, Popper would argue that what we are dealing with is a statistical
distribution which says nothing about the properties of individual
particles. However, we know from the study of electromagnetic theory that we
can explain single slit diffraction as a scattering involving currents in
the wall of the screen and the field in the slit. In this more complete
theory, the near field depends on the nature of the material composing the
screen, but, remarkably, the field a few wave lengths beyond the slit
primarily depends on the geometry of the slit. The same must be true for the
case of the diffraction of massive particles. In this regard, Heisenberg was
correct, for each individual particle such physical effects are present and
place physical limitations on how well we can determine the values of
conjugate variables. Note that Bohm's theory, which ascribes definite values
the position and momentum of each particle, does not allow us to measure the
state of a single particle any more precisely than quantum theory does. I do
agree with Popper that one should not call the passage of a particle through
a slit as a measurement but rather a selection.

\bibitem{Scully}  M. O. Scully and K. Dr\"{u}hl, Phys. Rev. A {\bf 25}, 2208
(1982).

\bibitem{Horne}  M. Horne, Fortschritte der Phys. {\bf 46}, 683 (1998).

\bibitem{Zeilinger}  A. Zeilinger, M. Horne, H. Weinfurter, and M. Zukowski,
Phys. Rev. Lett. {\bf 78}, 3031 (1997).

\bibitem{Bohmtext}  D. Bohm, {\it Quantum Theory}, (Prentice-Hall, Englewood
Cliffs, NJ, 1951).

\bibitem{Walther}  For a related discussion see M. O. Scully, B. Englert,
and H. Walther, Nature {\bf 351},111 (1991).

\bibitem{Bouwmeester}  D. Boouwmeester, {\it et. al.}, Nature {\bf 390}, 575
(1997).

\bibitem{Boschi}  D. Boschi, {\it et. al.}, Phys. Rev. Lett. {\bf 80}, 1121
(1998).
\end{references}
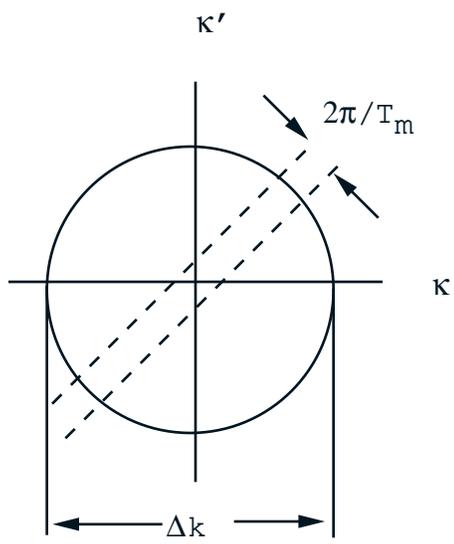
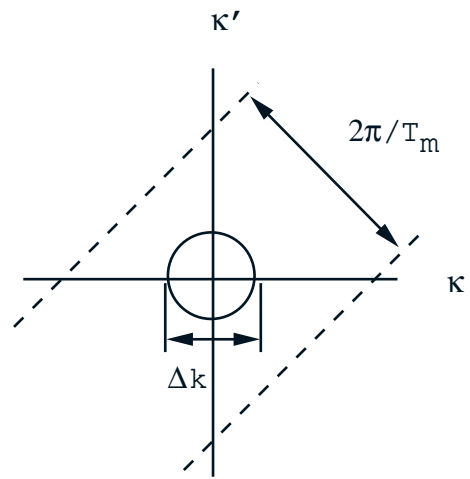
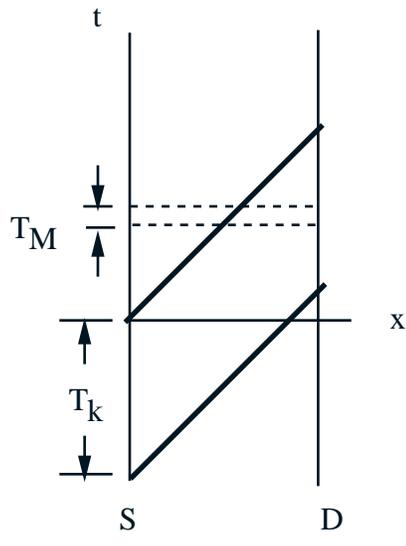
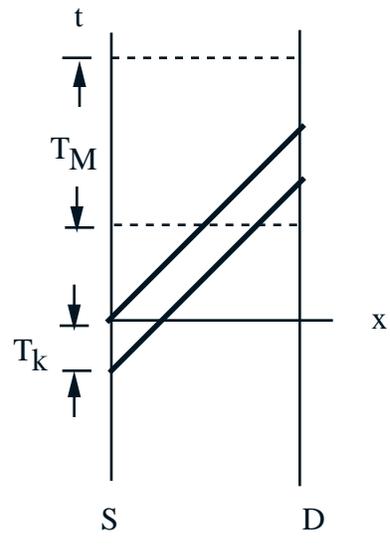
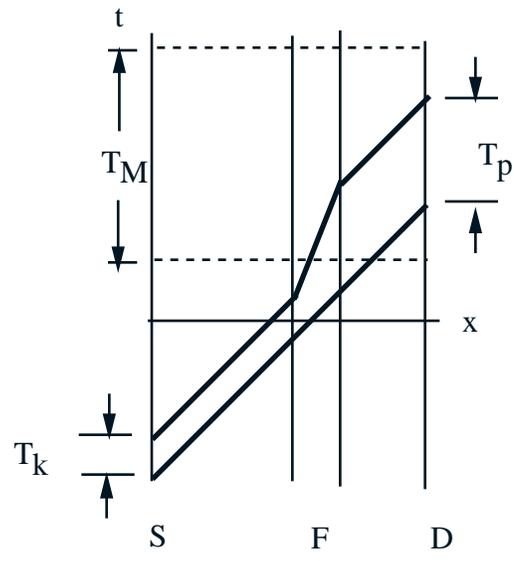
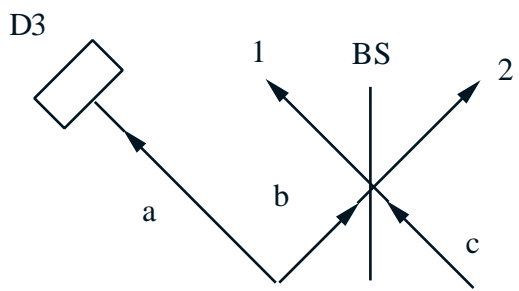
\end{document}